\documentclass[12pt]{article}
\usepackage{latexsym}
\usepackage{epsfig}

\newcommand{\beq}{\begin{eqnarray}}
\newcommand{\eeq}{\end{eqnarray}}
\newcommand{\be}{\begin{eqnarray*}}
\newcommand{\ee}{\end{eqnarray*}}
\newcommand{\bk}{{\bf k}}

\newcommand{\ra}{\rightarrow}

\newcommand{\ve}{\varepsilon}
\newcommand{\nn}{\nonumber}

\newcommand{\om}{{\omega}}

\begin{document}

\centerline{\Large\bf{A dispersive correction to the Casimir force}}
\bigskip
\centerline{Finn Ravndal$^a$ and Lee Peng Teo$^b$}
\bigskip
\centerline{$^a$\it Department of Physics, University of Oslo, Blindern, N-0316 Oslo, Norway.}
\medskip
\centerline{$^b$\it Department of Applied Mathematics, Faculty of Engineering,}
 \centerline{\it University of Nottingham Malaysia Campus, Jalan Broga,}
\centerline{\it 43500, Semenyih, Selangor Darul Ehsan, Malaysia.}

\begin{abstract}

\small{Using perturbation theory the first order dispersive correction to the Casimir energy between two plates separated by a dielectric material is calculated. It falls off with the plate separation as $1/L^6$. The result is derived both from evaluation of the zero-point energy and within the Lifshitz formulation. It is pointed out that a possible surface term can be more important, varying like $1/L^5$.}

\end{abstract}

The attractive Casimir force \cite{1} per unit area between two parallel, metallic plates separated by a distance $L$, has the well-known value
\beq
              F = -{\pi^2\hbar c\over 240 L^4}.                                           \label{vacuum}
\eeq
Today there exists a large literature around this quantum phenomenon \cite{2} and its experimental verifications \cite{3}.

In more realistic situations one can consider the same system with a dielectric medium between the plates instead of an ideal vacuum. At low energies, i.e. for long wavelengths, this medium can be considered to be continuous with a refractive index $n>1$ given by the square root of the dielectric constant. The resulting Casimir force in this case was calculated by Brevik and Milton using the Minkowski formalism for the electro-magnetic energy-momentum tensor and with field correlators  obtained from the fluctuation-dissipation theorem combined with more standard Green's functions methods \cite{BM}. After a rather lengthy calculation they obtained a force which was simply the ordinary vacuum result (\ref{vacuum}) reduced by the factor $n$. The same result was independently obtained by Teo \cite{Teo} based on a derivation similar  to the alternative Lifshitz theory \cite{Lif}.

Originally, the Casimir force was explained by the fluctuations of the electromagnetic field in the vacuum between the plates, giving rise to the zero-point energy $E = (1/2)\sum_\bk \hbar\om_\bk$ where $\bk$ denotes the wave vectors and $\om_\bk= c\bk$ are the corresponding frequencies. Due to the restricted geometry between the plates, the wave vectors are quantized and have typical magnitudes set by the separation $L$.  The velocity of light in the medium will have the reduced value $c \ra c/n$. Thus the zero-point energy is smaller by the same factor and therefore also the Casimir force. From this point of view \cite{casimir}, the result of Brevik and Milton then follows directly without any calculations.

All the above derivations of the material Casimir force assume that the electromagnetic waves in the medium suffer no dispersion, i.e. that the refractive index $n$ is independent of the frequency $\om$. Usually this dependency is very complex, depending on the detailed absorptive properties of the material. Since the dominant contributions to the Casimir force come from rather low frequencies set by the plate separation $L$ and assuming in the simplest case no such absorption, the dispersion is then described by the Cauchy formula \cite{Hecht}
\beq
              n(\om) = n_0 + n_1\om^2,                  \label{cauchy}
\eeq
where $n_0$ and $n_1$ are constants.

In order to calculate the effect of this dispersive correction, consider first the derivation of the Casimir energy when $n$ is constant. Using units so that $\hbar = 1$ and the   light velocity in vacuum $c=1$, the total zero-point energy per unit plate area is
\beq
            E_0 = \int{d^2k_T\over (2\pi)^2} \sum_{n=1}^\infty \om_{n\bk_T},            \label{sum}
\eeq
including the contribution from the two polarization directions. The quantized frequencies follow from the boundary conditions which give
\beq
              \om_{n\bk_T} =  {1\over n_0}\sqrt{{\bk_T}^2 + (n\pi/L)^2}.
\eeq
The sum-integral (\ref{sum}) is obviously divergent. For our purpose it is convenient to define it as the regularized value of
\be
            E_0 =  \int{d^2k_T\over (2\pi)^2} \sum_{n=1}^\infty \om_{n\bk_T} | \om_{n\bk_T}|^{-s}
\ee
in the limit $s\ra 0$. We then have
\beq
            E_0 &=&\lim_{s\ra 0} {n_0^{s-1}\over 4\pi} \sum_{n=1}^\infty \int_0^\infty \! dk_T^2 \left| k_T^2 +  (n\pi/L)^2\right|^{(1-s)/2} \nn \\
               &=& -\lim_{s\ra 0}{\pi^{2-s}n_0^{s-1}\over (6-2s)L^{3-s}}  \sum_{n=1}^\infty n^{3-s}  = - {\pi^2\over 6 n_0 L^3}\zeta(-3),   \label{lead}
\eeq
Since the needed value of the Riemann zeta-function is $\zeta(-3) =  1/120$, we have the finite result $E_0 = -\pi^2/720n_0 L^3$ for the zero-point energy between the plates. The force between them now follows from  $F = - dE/dL = -\pi^2/240n_0 L^4$ in agreement with the known result \cite{BM,Teo}.

The dispersive correction $\Delta E$ resulting from (\ref{cauchy}) will be small and can be derived  the same way  from (\ref{sum}) by the replacement $1/n_0 \ra 1/n(\om) = 1/n_0 - n_1\om^2/n_0^2$ to lowest order.  It gives
\be
                 \Delta E &=&  -{n_1\over n_0}\int{d^2k_T\over (2\pi)^2} \sum_{n=1}^\infty \om_{n\bk_T}^3 | \om_{n\bk_T}|^{-s}  \\
               &=& - \lim_{s\ra 0} {n_1\over 4\pi n_0^{4-s}} \sum_{n=1}^\infty \int_0^\infty \! dk_T^2 \left| k_T^2 +  (n\pi/L)^2\right|^{(3-s)/2}  \\
     &=&  \lim_{s\ra 0} {n_1\pi^{4-s}n_0^{s-4}\over (10-2s)L^{5-s}}    \sum_{n=1}^\infty n^{5-s} =  - {n_1\over  n_0^4} {\pi^4\over 2520 L^5},
\ee
since $\zeta(-5) =  -1/252$. Combining this with the above zero-order result, we have for the full fluctuation energy per unit plate area
\beq
                E = - {\pi^2\over 720 L^3n_0}\left(1 + {2\pi^2n_1\over7n_0^3L^2}\right).             \label{correct}
\eeq
The effect of the obtained correction is seen to be greatest at the smallest plate separations. In fact, from  (\ref{cauchy})  we know that the dispersive part  should satisfy  $n_1\om^2 < n_0$ in order to be treated as a perturbation.  Since $n_0 = {\cal O}(1)$ and the important frequencies are set by $\om = {\cal O}( 2\pi/L)$, we should have that $n_1/L^2 < 1/4\pi^2$.  Our result is therefore only valid for plate separations $L > 2\pi\sqrt{n_1}$. In this region the correction is therefore expected to satisfy $\Delta E/E_0 < 1/(14n_0^3)$.

Above we have derived the effect of dispersion in a somewhat phenomenological way. The zero-point energy is a consequence of the quantization of the electromagnetic field in the medium and it requires therefore a more fundamental formulation. This was recently suggested by the introduction of an effective Lagrangian for the electromagnetic field in a dielectric medium with dielectric constant $\ve$, assumed to be spatially constant \cite{eff}. It is obtained by adding higher-order operators to the free Lagrangian
\beq
              {\cal L}_0 = {1\over 2}\big(n^2{\bf E}^2 - {\bf B}^2\big),        \label{L_0}
\eeq
where the index of refraction $n  $ now is a constant. Both the field operators here have dimension four in the units we now use. The next possible operator has dimension six and is $\Delta {\cal L}_1 = a_1{\bf E}\cdot\nabla^2{\bf E}$ where the constant $a$ is set by the properties of the medium. It is now easy to show that when this is treated as an ordinary perturbation, its net effect is described by the dispersion relation (\ref{cauchy}). Among several possible operators with dimension eight, one can similarly show that $\Delta {\cal L}_2 = a_2(\nabla^2{\bf E})^2$ describes a dispersive correction $\propto \om^4$ which is smaller than what results from the previous operator. All these operators will modify the zero-point energy coming from (\ref{L_0}). The actual magnitudes of these corrections can be calculated by using standard methods from perturbative quantum field theory \cite{KR}. Needless to say, from the above dimension-six operator the lowest order correction (\ref{correct}) will thus result.

An alternative to this field theory description of the Casimir effect, one also has  the Lifshitz formalism \cite{Lif}.  Here the properties of the confining plates can be taken more directly into account. When these are ideal metals and the substance between them can be assigned an index of refraction $n(\om)$,  the Casimir energy is given by
\beq
              E = {1\over 2\pi^2}\int_0^\infty\! d\xi \int_{\kappa_1}^\infty\!d\kappa \kappa\log\big(1 - e^{-2\kappa L}\big),
\eeq
where the lower limit of integration is $\kappa_1 = {n(i\xi)\xi}$. With the Cauchy dispersion relation (\ref{cauchy}) it becomes $\kappa_1 = n_0\xi - n_1\xi^3$. The integral can therefore be split into two parts $E = E_0 + \Delta E$ with
\beq
           E_0 =   {1\over 2\pi^2}\int_0^\infty\! d\xi \int_{\kappa_0}^\infty\!d\kappa \kappa\log\big(1 - e^{-2\kappa L}\big),       \label{E_0}
\eeq
and
\beq
           \Delta E =   {1\over 2\pi^2}\int_0^\infty\! d\xi \int_{\kappa_1}^{\kappa_0}\!d\kappa \kappa\log\big(1 - e^{-2\kappa L}\big),   \label{DeltaE}
\eeq
where now $\kappa_0 = n_0\xi$. The first term represents the leading term (\ref{lead}). This follows from rewriting the double integration as
\be
           E_0 &=&  {1\over 2\pi^2}\int_0^\infty\! d\kappa\kappa \int_0^{\kappa/n_0}\!d\xi\log\big(1 - e^{-2\kappa L}\big)
          = {1\over 2\pi^2n_0}\int_0^\infty\! d\kappa\kappa^2\log\big(1 - e^{-2\kappa L}\big).
\ee
Expanding the logarithm and integrating, one then obtains
\beq
           E_0 = - {1\over 8\pi^2L^3n_0}\sum_{n=1}^\infty{1\over n^4} = -{\pi^2\over 720 L^3n_0},
\eeq
since the sum is given by $\zeta(4) = \pi^4/90$.

The correction (\ref{DeltaE}) can be similarly evaluated. To first order in the dispersive correction it becomes
\be
         \Delta E &=&   {1\over 2\pi^2}\int_0^\infty\! d\xi (\kappa_0 - \kappa_1)\kappa_0\log\big(1 - e^{-2\kappa_0L}\big) \\
            &=& {n_1n_0\over 2\pi^2}\int_0^\infty\! d\xi \xi^4 \log\big(1 - e^{-2n_0\xi L}\big) =  {2\pi^2n_1\over7n_0^3L^2} E_0,
\ee
where in the integral we have used $\zeta(6) = \pi^6/945$. We thus find the same total Casimir energy (\ref{correct}) as obtained directly from the zero-point energy between the plates.

In addition to this dispersive correction, one also  expects higher-order quantum effects to modify  the result as for the Casimir energy in vacuum \cite{Bordag}. These were later shown to be described by a surface term in the effective action \cite{RT}. It corresponds to the free Lagrangian (\ref{L_0}) constrained to the plates and reduced by an effective coupling constant. Since it appears in one spatial dimension less, its net effect to the Casimir energy will be similar to a dimension-five bulk operator and thus vary with the separation like $1/L^5$. Its overall magnitude can be calculated in vacuum \cite{Cliff}, but for the interface between a dielectric medium and a metal plate, it's not clear what the magnitude of the corresponding coupling constant will be. If it is sufficiently large, it will dominate over the dispersive correction calculated here.


\begin{thebibliography}{99}

\bibitem{1} H. B. G. Casimir, Proc. K. Ned. Akad. Wet.  {\bf 51} 793 (1948).

\bibitem{2} P. W. Milonni, {\it The Quantum Vacuum}, Academic Press, San Diego (1994); V. M. Mostepanenko, N. N. Trunov, {\it The Casimir effect and its applications}, Clarendon Press, Oxford (1997); K. A. Milton, {\it The Casimir effect}, World Scientific, Singapore  (2001);  M. Bordag, G.L. Klimchitskaya, U. Mohideen and V. M. Mostepanenko, {\it Advances in the Casimir effect}, Oxford University Press, Oxford (2009).

\bibitem{3}  S. K. Lamoreaux, {\it Rep. Prog. Phys.} {\bf 68}, 201 (2005).

\bibitem{BM}  I. Brevik and K.A. Milton,  {\it Phys. Rev.} {\bf E78},  011124  (2008).

\bibitem{Teo} L.P. Teo,  {\it Phys. Rev.} {\bf A81},  032502  (2010).

\bibitem{Lif}  E.M. Lifshitz, {\it Zh. Eksp. Teor. Fiz}, {\bf 29}, 94 (1956);   L.D. Landau and E.M. Lifshitz, {\it Statistical Physics II}, Pergamon Press, Oxford (1980); M. Bordag, U. Mohideen and V. M. Mostepanenko, {\it Phys. Rep.} {\bf 353}, 1 (2001).

\bibitem{casimir} F. Ravndal, {\it Phys. Rev.} {\bf E79}, 053101 (2009).

\bibitem{Hecht} E. Hecht, {\it Optics}, Addison-Wesley, New York (1998).

\bibitem{eff}  F. Ravndal,   arXiv:quant-ph/0804.4013;  arXiv:hep-ph/0810.1872.

\bibitem{KR} X. Kong and F. Ravndal, {\it Phys. Rev. Lett.} {\bf 79}, 545 (1997).

\bibitem{Bordag}  M. Bordag, D. Robaschik and E. Wieczorek, {\it Ann. Phys.} {\bf 165}, 192 (1985).

\bibitem{RT} F. Ravndal and J.B. Thomassen, {\it Phys. Rev.} {\bf D63}, 113007 (2001).

\bibitem{Cliff}  Y. Aghababaie and C.P. Burgess, {\it Phys. Rev.}  {\bf D70}, 085003 (2004).

\end{thebibliography}
\end{document}